# Magnetic properties of submicron Co islands and their use as artificial pinning centers

M. J. Van Bael, K. Temst, V. V. Moshchalkov, and Y. Bruynseraede

*Laboratorium voor Vaste-Stoffysica en Magnetisme, Katholieke Universiteit Leuven, Celestijnenlaan 200D, B-3001 Leuven, Belgium*

We report on the magnetic properties of elongated submicron magnetic islands and their influence on a superconducting film. The magnetic properties were studied by magnetization hysteresis loop measurements and scanning-force microscopy. In the as-grown state, the islands have a magnetic structure consisting of two antiparallel domains. This stable domain configuration has been directly visualized as a $2\times2$-checkerboard pattern by magnetic-force microscopy. In the remanent state, after magnetic saturation along the easy axis, all islands have a single-domain structure with the magnetic moment oriented along the magnetizing field direction. Periodic lattices of these Co islands act as efficient artificial pinning arrays for the flux lines in a superconducting Pb film deposited on top of the Co islands. The influence of the magnetic state of the dots on their pinning efficiency is investigated in these films, before and after the Co dots are magnetized.
[S0163-1829(99)13221-1]

## I. INTRODUCTION

Recent advances in magnetic device and storage technology have strongly stimulated the research efforts focused on nanostructured magnetic films.[1–14] An important aspect of this research is related to the development of nonvolatile magnetic memories, based on submicron magnetic dots.[10,15] Such single-bit-per-particle memories consist of large arrays of noninteracting magnetic particles,[3,14] which are ideally in a single-domain state with two possible stable configurations to be associated with the logical ''1'' and ''0.'' To impose these two stable states, one can make use of the shape anisotropy, since the precise shape and dimensions of the islands are determining factors for their magnetic properties.[2,6,10–13,16] Particularly submicron elongated or ellipsoidal magnetic islands have a clearly defined direction along which the magnetization **M** can be oriented.[8,10,13,16] Huge two-dimensional (2D) arrays of magnetic dots are also used in other research fields, e.g., to create a periodic spatial modulation of the magnetic field in a 2D electron gas to study Weiss oscillations of the magnetoresistance,[17–22] or as an artificial pinning array in combination with a superconducting film to investigate the pinning of flux lines.[23–25] Many interesting phenomena have been observed in superconducting films and multilayers with a periodic lattice of *artificial pinning centers*, such as well-controlled defects[26] or submicron holes (antidots).[27–30] The interplay between the regular pinning array and the periodic vortex lattice gives rise to pronounced commensurability effects in the critical current density $j_c$ as function of the perpendicular applied magnetic field $H$. Recently, Martin *et al.* have successfully used a lattice of *ferromagnetic dots* (Fe or Ni) to create a periodic array of artificial pinning centers in a superconducting Nb film.[23] Clear matching effects are observed in the $j_c(H)$ behavior of these systems.

In this paper, we report on the study of submicron rectangular Au/Co/Au islands with in-plane magnetization. Large triangular and square lattices, covering an area of approximately 6 mm$^2$ and containing $\sim 10^6$ noninteracting islands, are prepared by liftoff lithographic techniques. The structure is characterized by atomic-force microscopy (AFM), the magnetic properties are studied by superconducting quantum interference device (SQUID) magnetization measurements. In order to obtain local microscopic information on the domain structure of the individual islands, magnetic-force microscopy (MFM) measurements have been performed. We will also demonstrate how these periodic lattices of magnetic dots act as artificial pinning arrays for the flux lines in a superconducting film. A periodic pinning potential is created in a superconducting Pb film that is grown on top of the magnetic island lattices. The pinning properties and matching effects are studied by means of SQUID magnetization measurements and the influence of the stray field of the magnetic dots on their pinning properties is demonstrated.

## II. EXPERIMENT

The island lattices are prepared using electron-beam lithography and liftoff techniques. A Au(75 Å)/Co(200 Å)/Au(75 Å) trilayer is deposited on a SiO$_2$ substrate covered by a double resist layer in which a triangular or a square lattice of submicron rectangular holes was defined by electron-beam lithography. The double resist layer results in an overhang resist profile avoiding a connection of the deposited material inside the holes with material remaining on top of the resist. The layers are grown at room temperature in a molecular-beam epitaxy system at a working pressure of $2\times10^{-10}$ Torr. The Au layers are evaporated from a Knudsen cell at a rate of 0.45 Å/s, whereas for the Co layer electron-beam evaporation was applied at a rate of 0.25 Å/s. The evaporation rates are controlled using a quadrupole mass spectrometer. After the deposition of the layers, the remaining resist was removed in hot acetone, leaving a lattice of Au/Co/Au islands on the substrate. The Au buffer and top layers are necessary to obtain islands with clean edges after liftoff. Grown in the above-mentioned conditions, the samples have a polycrystalline structure, as was confirmed by x-ray-diffraction experiments.

The AFM and MFM experiments are performed using a Digital Instruments Nanoscope III system. The magnetic tip in the MFM has a magnetic moment oriented perpendicular to the scanning plane, sensitive to the perpendicular component of the magnetic stray field extruding from the sample surface.[31–33] The MFM experiments are performed in the ac mode using the phase detection technique.[33] The topographical features are separated from the magnetic signal by scanning each scan line twice in a two-step lift mode.

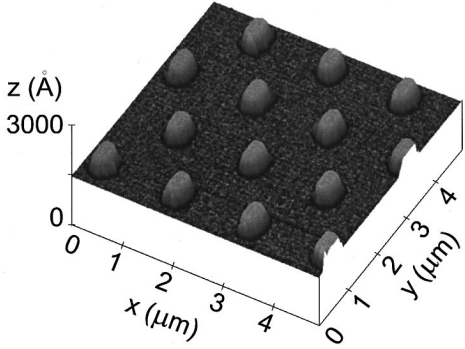

FIG. 1. AFM image of a $(5 \times 5)$ $\mu m^2$ area of a triangular lattice of elongated Au(75 Å)/Co(200 Å)/Au(75 Å) islands with lattice period 1.5 $\mu m$.

## III. MAGNETIC PROPERTIES

The shape and size of the islands are examined by AFM measurements. Figure 1 shows an AFM topograph of a $(5 \times 5)$ $\mu m^2$ area of a triangular lattice with a period of 1.5 $\mu m$. The islands have a smooth upper surface; their shape is rectangular with rounded corners. The lateral dimensions of the islands are determined from a section analysis of the AFM topographs and are $L_s = (3600 \pm 100)$ Å and $L_l = (5400 \pm 100)$ Å, resulting in an aspect ratio of 3:2. More detailed AFM measurements have shown that the shape and size of the islands are homogeneous over the full lattice.

Magnetic hysteresis loops of the island lattice are measured in a Quantum Design SQUID magnetometer in order to obtain information on the macroscopic magnetic behavior of the whole island lattice. For the lattice structure, as well as for a nonpatterned reference Au(75 Å)/Co(200 Å)/Au(75 Å) film grown in exactly the same conditions, the easy magnetization axis lies in the substrate plane. From hysteresis loop measurements of the reference film with the magnetic field applied in the film plane we have determined the saturation field $H_s = 600$ Oe and the coercive field $H_c = 160$ Oe. The hysteresis loops of the square and triangular island lattices show an additional in-plane anisotropy due to the lateral patterning. The results for the square lattice, measured at 5 K with the field oriented in the substrate plane both along the long ($L_l$) and the short ($L_s$) direction of the islands, are shown in Figs. 2(a) and 2(b) respectively. The linear diamagnetic background signal from the substrate has been subtracted. The slope of this background was determined by measuring the magnetization in a field region well above the saturation field for the island lattice. In case the field is oriented along the long direction of the islands, $\mathbf{H} \| \mathbf{L}_l$ [Fig. 2(a)], we find that $H_s = 750$ Oe and $H_c = 300$ Oe. When the field is applied along the short side of the islands, $\mathbf{H} \| \mathbf{L}_s$ [Fig. 2(b)], we observe that $H_s = 1250$ Oe and $H_c = 130$ Oe. This indicates that the easy axis for magnetization is along the

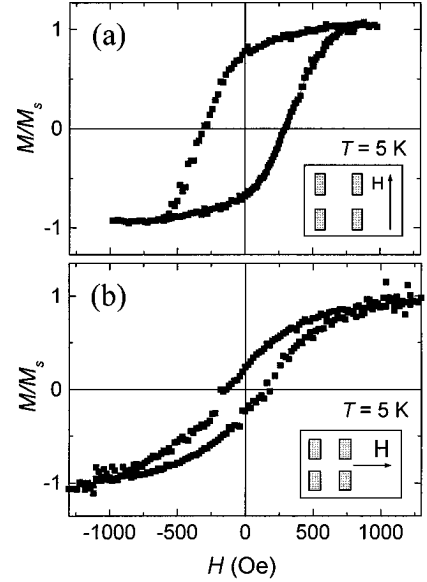

FIG. 2. Macroscopic hysteresis loops $M/M_s(H)$ with $M_s$ the saturation magnetization, measured at $T = 5$ K, of a square lattice of elongated Au(75 Å)/Co(200 Å)/Au(75 Å) islands. The field is oriented in the substrate plane (a) parallel to the long side of the elongated islands ($\mathbf{H} \| \mathbf{L}_l$), and (b) along the short side of the islands ($\mathbf{H} \| \mathbf{L}_s$). The field orientation with respect to the islands is schematically presented in the inset.

long direction of the islands, as can be expected from the shape anisotropy. The same results are also obtained for the triangular lattice of Au/Co/Au islands. The coercivity of the islands along the easy axis is increased as compared to the reference film, and is expected to increase further with increasing aspect ratio.[2,9] Similar magnetization measurements at 250 K, i.e., close to room temperature where the MFM measurements are performed, show the same results. This is not surprising since this temperature range is well below the Curie temperature of Co (1388 K).

The local magnetic domain structure of the individual islands was examined by means of MFM at room temperature and in zero field. The typical results of the MFM measurements are shown in Figs. 3(a) and 3(d) displaying six islands in a $(2.7 \times 4)$ $\mu m^2$ area of a square island lattice. The magnetic contrast originates from the component of the magnetic-force gradient perpendicular to the substrate plane. The white and black regions can thus be interpreted as the north and south poles of in-plane oriented magnetic domains, respectively, where perpendicular components arise due to the closing of the magnetic-field lines.[33]

Figure 3(a) shows a MFM image of the as-grown sample, i.e., before the application of an external field. Each of the six displayed islands shows two bright and two dark spots arranged in a $2 \times 2$-checkerboard configuration. This image remained stable while scanning it several times and in different scan directions. From larger scale MFM images, it is clear that the majority of the islands show this peculiar magnetic state. The two possible arrangements (the first one with the bright spots above-left and below-right, the second one corresponding to its mirror image) occur in an approximately 1 to 1 proportion without any long-range ordering. This indicates that there is no interaction between neighboring is-lands and, hence, that they behave independently. The magnetic image before magnetization of the islands can be interpreted as due to the fact that each island consists of *two antiparallel magnetic domains*.[34] Since the islands are too large to be spontaneously in a single magnetic domain state,[8,9] they will form two or more smaller magnetic domains, in a way that is strongly affected by their specific structural and morphological properties. Splitting into two *antiparallel* domains results in a zero total magnetic moment and thus substantially reduces the energy related to the existence of large stray fields. Creating domain walls also requires energy, and therefore we expect that the islands will find the state with the smallest possible area of the domain wall. This corresponds to a configuration of magnetic domains where the magnetic moments of the two domains are aligned along the short direction of the islands, $M \| L_s$ [see Fig. 3(b)]. However, since the domain state with two antiparallel domains directed along the long island axis $L_l$ [see Fig. 3(c)] would create qualitatively the same pattern in the MFM measurements, no decisive answer can be given concerning the orientation of the domains in the as-grown islands.

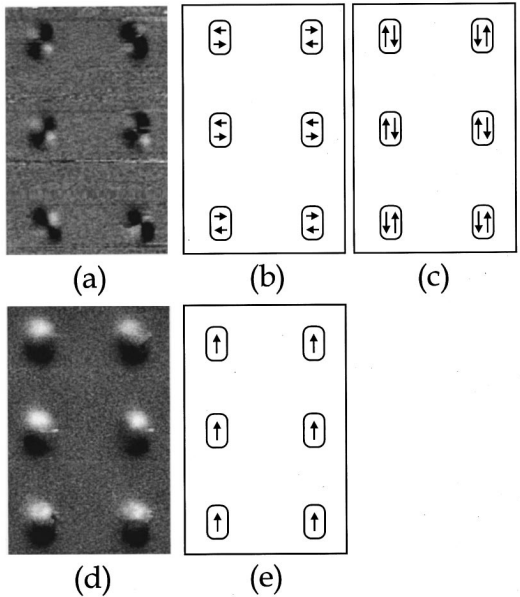

FIG. 3. (a) and (d) MFM images recorded at room temperature in zero field of a (2.7×4.0) $\mu m^2$ area of the island lattice, containing six magnetic islands. (a) shows the magnetic response of the islands before applying a magnetic field. The magnetic image of a similar area of the sample in the remanent state, after magnetic saturation in a magnetic field of 10 kOe pointing upwards, is given in (d). The suggested remanent magnetic domain configurations of the islands before and after magnetization are schematically presented in (b) and (c), and in (e), respectively.

Figure 3(d) shows the MFM image of a similar region of the same sample in the remanent state, after being magnetized outside the microscope in an external field of 10 kOe along the easy axis $L_l$. The image of every island consists of a bright spot and a dark spot, which is typical for a single magnetic domain particle with in-plane magnetization $M \| L_l$. This indicates that after being magnetized along the easy direction, the islands in the remanent state ($H=0$) remain in a state similar to the saturated one, where all magnetic islands are lined up by the applied field. The magnetic domain state of the islands after magnetization is schematically presented in Fig. 3(e). Larger area MFM scans of the remanent state show that this is the case for more than 85% of the islands. Similar observations have been reported for elongated submicron islands of different materials and with different geometrical properties.[3,8,9] Note that the term ''single domain'' may only be considered here within the resolution of the MFM, which is of the order of a few hundred Å. Possible disturbances in the alignment of the magnetic moments near the island edges and minor domains may be screened out by the overall magnetization of the island, and may therefore not be detectable by the MFM tip. Comparison of the SQUID magnetization measurements [Fig. 2(a)] with the MFM results indeed shows that the magnetic state in the remanence after magnetization can not be exactly the same as in saturation. This becomes clear by considering that a changed magnetization in less than 15% of the islands can not account for a 50% decrease of the magnetic moment of the remanent state with respect to the saturation magnetization [see Fig. 2(a)].

Based on these observations, we can identify three stable magnetic states for the studied elongated Au/Co/Au islands: (i) The state with two antiparallel magnetic domains in the as-grown islands, (ii) the remanent single-domain state with a positive magnetic moment, obtained after magnetic saturation in a positive field, and (iii) the remanent single-domain state with a negative total moment after magnetic saturation in a negative field.

## IV. Au/Co/Au-ISLAND LATTICE AS ARTIFICIAL PINNING ARRAY

It has been shown recently that a lattice of submicron ferromagnetic Fe or Ni dots creates a strong periodic pinning potential for the flux lines in a superconducting Nb film that is grown on top of it.[23] In the following experiment, we will study to what extent a periodic lattice of the above discussed magnetic Au/Co/Au dots can inhibit the vortex motion and increase the critical current density $j_c$ in a continuous superconducting Pb film. For that purpose, a 500 Å Pb film is deposited on top of a triangular island lattice by means of electron-beam evaporation at 77 K, using an evaporation rate of 9 Å/s. The sample is finally covered with a 200 Å protecting Ge layer.

SQUID magnetization measurements are performed on a Pb film with a triangular lattice of Au/Co/Au dots, before and after magnetizing the dots. It can be shown that the width of the magnetization loops, $\Delta M(H)$, is proportional to the superconducting critical current density $j_c(H)$.[35] The *magnetized* dots exhibit a single-domain structure [Fig. 3(d)], which is associated with substantially larger stray fields compared to the multidomain state before magnetizing [Fig. 3(a)]. These stray fields have a destructive effect on superconductivity and create a periodic array of areas with weakened superconducting properties or even normal-state areas, if the stray field is comparable to the upper critical field of the superconducting film. The possibility to change the magnetic state of the islands enables us to study *the influence of the magnetic stray field* of the islands on the pinning effects.

Figure 4 shows the upper branch of the magnetization loops, measured at $T/T_c=0.97$ ($T_c\approx 7.2$ K is the superconducting critical temperature) for a 500 Å Pb film on top of a triangular lattice of Au/Co/Au islands (before and after magnetizing the islands) and for a reference 500 Å Pb film. The magnetization is shown as function of $H/H_1$, where $H_1 = 2/\sqrt{3}[\phi_0/(1.5\,\mu\text{m})^2] = 10.6$ Oe (with $\phi_0$ the flux quantum), represents the first matching field for which the density of flux lines equals the density of dots. This implies that a one-to-one matching of the triangular vortex lattice onto the pinning array can be established at the first matching field $H_1$. The presence of the lattice of Au/Co/Au dots results in a very strong enhancement of the width of the magnetization loop $\Delta M \sim j_c$, compared to the reference Pb (500 Å) film without dots (line). Moreover, pronounced anomalies are observed for $T$ close to $T_c$ at certain multiples of $H_1$, e.g., a sudden drop of $M(H)$ at $H/H_1=1$ and a maximum at $H/H_1=4$ and, after the dots are magnetized, also at $H/H_1=2$ and 3. These matching effects indicate that the lattice of Au/Co/Au dots creates a strong periodic pinning potential for the flux lines in the Pb film that is deposited on top of it, similar to thin films or multilayers with an antidot lattice.[27–29]

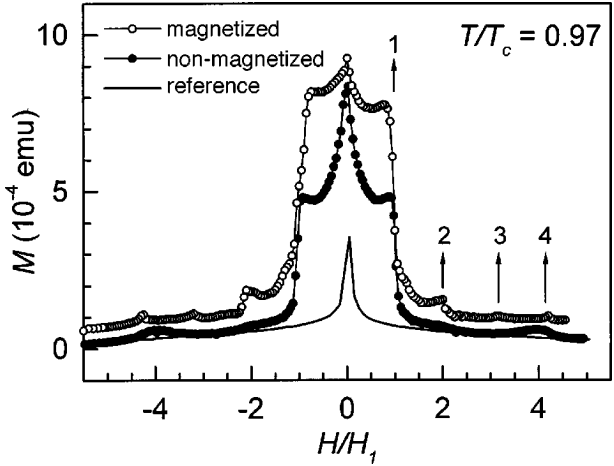
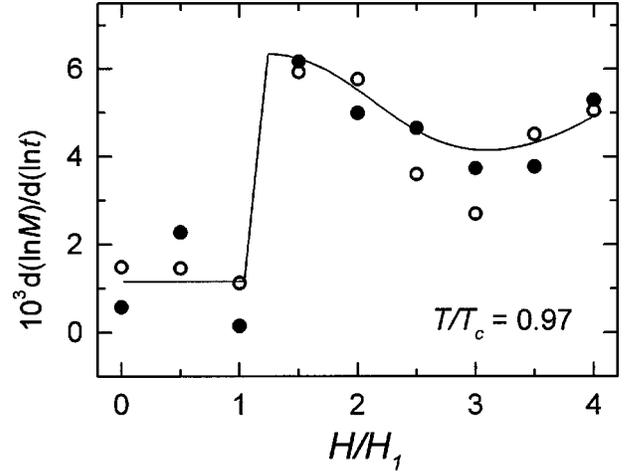

FIG. 4. Upper half of the magnetization loop $M$ vs $H/H_1$ at $T/T_c=0.97$, for a 500 Å Pb film on a triangular lattice of Au/Co/Au dots (period 1.5 $\mu$m) before (filled symbols) and after (open symbols) magnetizing the dots, and for a reference Pb (500 Å) film (line). For clarity, the curve for the magnetized dots is slightly shifted upwards.

FIG. 5. Normalized relaxation rate $d\ln(M(\text{emu}))/d\ln(t(s))$ as function of $H/H_1$ at $T/T_c=0.97$ for a 500 Å Pb film on a triangular lattice (period 1.5 $\mu$m) of nonmagnetized (filled symbols) and magnetized (open symbols) Au/Co/Au dots. The solid line is a guide to the eye.

All observed matching anomalies are associated with the formation of specific stable vortex configurations in the imposed triangular pinning potential. In order to identify them, it is important to note that only one flux quantum can be pinned at each dot. This can be deduced from the shape of the $M(H)$ loops, where the sudden decrease of $\Delta M$ at $H_1$ (Fig. 4) indicates a strong increase of the vortex mobility at $H>H_1$.[27,36,37] This can be explained by the existence of *interstitial* vortices for $H>H_1$, which are at positions between the dots, not pinned at the dots, and are therefore characterized by a higher mobility. Direct information on the vortex mobility can be obtained by measuring the magnetization relaxation rate. Figure 5 shows the normalized relaxation rate $[d(\ln M)/d(\ln t)$ with $t$ the time] versus $H/H_1$ measured in the remanent magnetization regime at $T/T_c=0.97$, before and after magnetizing the islands. In both cases, a sudden increase of the mobility at $H=H_1$ can be associated with the appearance of interstitial vortices and confirms that not more than one single flux quantum can be trapped at each island.[27,36,37] In Fig. 6, we propose the geometry of the vortex lattices that are stabilized at those multiples of $H_1$ for which a matching anomaly is observed. At $H_1$, the vortex lattice identically matches the triangular pinning array. For $H>H_1$, interstitial vortices are ''caged'' between the strongly pinned vortices at the pinning sites. At $H/H_1=3$ and 4, a commensurate triangular vortex lattice with a smaller unit cell matches the triangular pinning potential, and is rotated over 30° in the case $H/H_1=3$. At $H/H_1=2$, a honeycomb vortex configuration is suggested, consisting of two identical interpenetrating triangular vortex lattices, both with the same period as the island lattice (1.5 $\mu$m). The suggested configurations at $H/H_1=1$, 3, and 4 have also been found by Reichhardt *et al.* in molecular-dynamics simulations of the vortex lattice in a triangular array of small pinning centers.[38] The ordered vortex configuration at $H/H_1 = 2$ is however only found in these simulations provided that the pinning at the pinning centers is sufficiently strong.

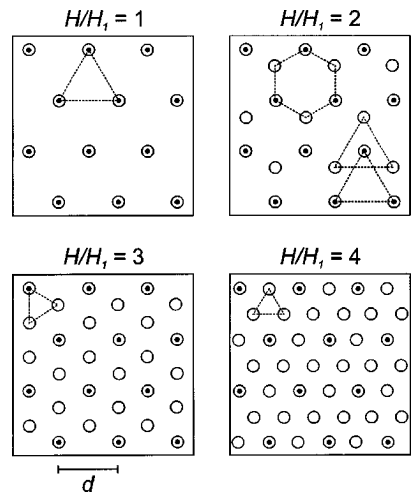

FIG. 6. Schematic presentation of the stable vortex configurations at integer matching fields for a triangular lattice of pinning centers with period $d$. Black dots and open circles represent pinning centers and vortices, respectively. Dashed lines indicate the symmetry of the stabilized vortex lattice.

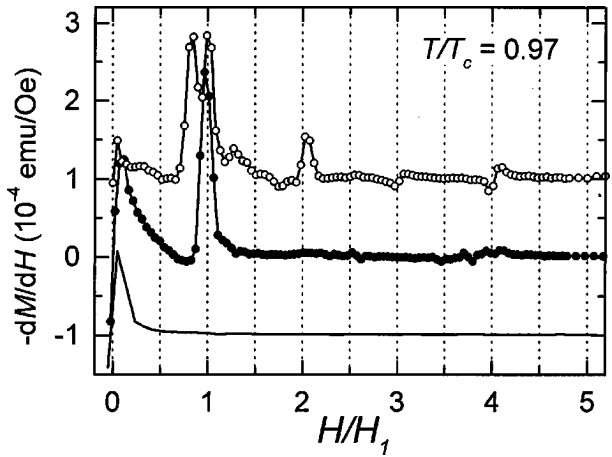

FIG. 7. Derivative $-dM/dH$ vs $H/H_1$ at $T/T_c = 0.97$ for a 500 Å Pb film on a triangular lattice of Au/Co/Au dots (period 1.5 μm) and for a reference 500 Å Pb film (line). The curves for the nonmagnetized and magnetized dots are represented by filled and open symbols, respectively, and are vertically displaced for clarity.

In contrast to the results of Martin *et al.*,[23] who found no influence of the magnetic state of the dots on their pinning properties, our experiments indicate that the stray field of the dots plays an important role. A clear influence of the stray field of the Au/Co/Au islands on the pinning properties can be noticed when comparing the magnetization curves in Fig. 4 for the magnetized (open symbols) and the nonmagnetized islands (filled symbols). It can be seen that after magnetizing the islands (i) $\Delta M$ ($\sim j_c$) is further increased and (ii) additional matching anomalies appear at $H/H_1 = 2$ and 3, and the anomaly at $H/H_1 = 4$ is sharper.

In Fig. 7, the differential signal $-dM/dH$ is plotted as function of $H/H_1$, where the filled and the open symbols represent the nonmagnetized and the magnetized islands, respectively. After magnetizing the dots, the appearance of additional peaks at $H/H_1 = 2$ and 3 and a more pronounced anomaly at $H/H_1 = 4$ can be clearly observed. The behavior at the second matching field is most prominent and demonstrates that pinning is stronger after the islands are magnetized. Indeed, molecular-dynamics simulations have shown that for a triangular lattice of pinning centers which can only trap one flux quantum, an anomaly at $H/H_1 = 2$ is only expected for sufficiently strong pinning centers.[38] Combining this result with our observation that a peak at $H/H_1 = 2$ is only observed *after* magnetizing the islands, leads to the conclusion that the stronger stray field of the dots enhances the pinning efficiency, and favors the formation of composite flux lattices.

## V. CONCLUSION

In conclusion, we have observed a spontaneous appearance of a magnetic configuration with two antiparallel magnetic domains in elongated magnetic islands. In the remanent state after magnetization, the islands remain in a single-domain state with their magnetic moment aligned with the imposed magnetizing field.

Triangular lattices of these magnetic dots act as an efficient artificial pinning array for the flux lines in a superconducting Pb film in a perpendicular field. Our results show that the stray field of the magnetic islands plays an important role in their pinning properties, and indicate that the pinning is stronger after the islands are magnetized.

## ACKNOWLEDGMENTS

The authors would like to thank Chris Van Haesendonck and Eddy Kunnen for fruitful discussion and Rik Jonckheere for the preparation of the resist patterns. This work was supported by the Belgian Inter-University Attraction Poles (IUAP) and Concerted Research Actions (GOA) programs and by the Fund for Scientific Research-Flanders (FWO). M.J.V.B. and K.T. would like to thank the FWO for financial support.

[1] J. F. Smyth, S. Schultz, D. Kern, H. Schmid, and D. Yee, J. Appl. Phys. **63**, 4237 (1988).

[2] J. F. Smyth, S. Schultz, D. R. Fredkin, D. P. Kern, S. A. Rishton, H. Schmid, M. Cali, and T. R. Koehler, J. Appl. Phys. **69**, 5262 (1991).

[3] G. A. Gibson and S. Schultz, J. Appl. Phys. **73**, 4516 (1993).

[4] C. Shearwood, S. J. Blundell, M. J. Baird, J. A. C. Bland, M. Gester, H. Ahmed, and H. P. Hughes, J. Appl. Phys. **75**, 5249 (1994).

[5] A. D. Kent, S. von Molnar, S. Gider, and D. D. Awschalom, J. Appl. Phys. **76**, 6656 (1994).

[6] F. Rousseaux, D. Decanini, F. Carcenac, E. Cambril, M. F. Ravet, C. Chappert, N. Bardou, B. Bartenlian, and P. Veillet, J. Vac. Sci. Technol. B **13**, 2787 (1995).

[7] P. R. Krauss and S. Y. Chou, J. Vac. Sci. Technol. B **13**, 2850 (1995).

[8] R. M. H. New, R. F. W. Pease, and R. L. White, J. Vac. Sci. Technol. B **13**, 1089 (1995).

[9] R. D. Gomez, M. C. Shih, R. M. H. New, R. F. W. Pease, and R. L. White, J. Appl. Phys. **80**, 342 (1996).

[10] S. Y. Chou, P. R. Krauss, and L. Kong, J. Appl. Phys. **79**, 6101 (1996).

[11] M. Hehn, K. Ounadjela, J.-P. Bucher, F. Rousseaux, D. Decanini, B. Bartenlian, and C. Chappert, Science **272**, 1782 (1996).

[12] A. O. Adeyeye, J. A. C. Bland, C. Daboo, Jaeyong Lee, U. Ebels, and H. Ahmed, J. Appl. Phys. **79**, 6120 (1996).

[13] A. O. Adeyeye, G. Lauhoff, J. A. C. Bland, C. Daboo, D. G. Hasko, and H. Ahmed, Appl. Phys. Lett. **70**, 1046 (1997).

[14] A. O. Adeyeye, J. A. C. Bland, C. Daboo, and D. G. Hasko, Phys. Rev. B **56**, 3265 (1997).

[15] R. M. H. New, R. F. W. Pease, and R. L. White, J. Vac. Sci. Technol. B **12**, 3196 (1994).

[16] M. S. Wei and S. Y. Chou, J. Appl. Phys. **76**, 6679 (1994).
[17] D. Weiss, K. von Klitzing, K. Ploog, and G. Weimann, Europhys. Lett. **8**, 179 (1989).
[18] R. R. Gerhardts, D. Weiss, and K. von Klitzing, Phys. Rev. Lett. **62**, 1173 (1989).
[19] F. M. Peeters and P. Vasilopoulos, Phys. Rev. B **47**, 1466 (1993).
[20] P. D. Ye, D. Weiss, R. R. Gerhardts, M. Seeger, K. von Klitzing, K. Eberl, and H. Nickel, Phys. Rev. Lett. **74**, 3013 (1995).
[21] R. R. Gerhardts, Phys. Rev. B **53**, 11 064 (1996).
[22] P. D. Ye, D. Weiss, R. R. Gerhardts, and H. Nickel, J. Appl. Phys. **81**, 5444 (1997).
[23] J. I. Martín, M. Vélez, J. Nogués, and Ivan K. Schuller, Phys. Rev. Lett. **79**, 1929 (1997).
[24] Y. Nozaki, Y. Otani, K. Runge, H. Miyajima, B. Pannetier, J. P. Nozières, and G. Fillion, J. Appl. Phys. **79**, 8571 (1996).
[25] O. Geoffroy, D. Givord, Y. Otani, B. Pannetier, and F. Ossart, J. Magn. Magn. Mater. **121**, 223 (1993).
[26] K. Harada, O. Kamimura, H. Kasai, T. Matsuda, A. Tonomura, and V. V. Moshchalkov, Science **274**, 1167 (1996).
[27] M. Baert, V. V. Metlushko, R. Jonckheere, V. V. Moshchalkov, and Y. Bruynseraede, Phys. Rev. Lett. **74**, 3269 (1995).
[28] V. V. Moshchalkov, M. Baert, V. V. Metlushko, E. Rosseel, M. J. Van Bael, K. Temst, R. Jonckheere, and Y. Bruynseraede, Phys. Rev. B **54**, 7385 (1996).
[29] V. V. Moshchalkov, M. Baert, V. V. Metlushko, E. Rosseel, M. J. Van Bael, K. Temst, Y. Bruynseraede, and R. Jonckheere, Phys. Rev. B **57**, 3615 (1998).
[30] A. Castellanos, R. Wördenweber, G. Ockenfuss, A. v. d. Hart, and K. Keck, Appl. Phys. Lett. **71**, 962 (1997).
[31] Y. Martin and H. K. Wickramasinghe, Appl. Phys. Lett. **50**, 1455 (1987).
[32] J. J. Saenz, N. Garcia, P. Grutter, E. Meyer, H. Heinzelmann, R. Wiesendanger, L. Rosenthaler, H. R. Hidber, and H.-J. Güntherodt, J. Appl. Phys. **62**, 4293 (1987).
[33] *Scanning Tunneling Microscopy II*, edited by R. Wiesendanger and H.-J. Güntherodt (Springer-Verlag, Berlin, 1992).
[34] E. F. Wassermann, M. Thielen, S. Kirsch, A. Pollmann, H. Weinforth, and A. Carl, J. Appl. Phys. **83**, 1753 (1998).
[35] C. P. Bean, Phys. Rev. Lett. **8**, 250 (1962).
[36] E. Rosseel, M. J. Van Bael, M. Baert, R. Jonckheere, V. V. Moshchalkov, and Y. Bruynseraede, Phys. Rev. B **53**, R2983 (1996).
[37] K. M. Beauchamp, T. F. Rosenbaum, U. Welp, G. W. Crabtree, and V. M. Vinokur, Phys. Rev. Lett. **75**, 3942 (1995).
[38] C. Reichhardt, C. J. Olson, and F. Nori, Phys. Rev. B **57**, 7937 (1998).